\documentclass[aps,prc,twocolumn,times,graphicx,tighten,showpacs]{revtex4}
\usepackage[dvips]{graphicx}
\usepackage[dvips]{color}

\newcommand{\bea}{\begin{eqnarray}}
\newcommand{\eea}{\end{eqnarray}}
\newcommand{\be}{\begin{equation}}
\newcommand{\ee}{\end{equation}}
\newcommand{\np}{{\bf p}}
\newcommand{\nh}{{\bf h}}
\newcommand{\nq}{{\bf q}}
\newcommand{\Qbar}{\not{\!Q}}
\newcommand{\Kbar}{\not{\!K}}
\newcommand{\Pbar}{\not{\!P}}
\newcommand{\tauvec}{\mbox{\boldmath $\tau$}}

\begin{document}

\title{
Pionic correlations and meson-exchange currents in two-particle
emission induced by electron scattering }

\author{
J.E. Amaro$^a$,
C. Maieron$^a$,
M.B. Barbaro$^b$,
J.A. Caballero$^c$,
T.W. Donnelly$^d$,
}

\affiliation{$^a$Departamento de F\'{\i}sica At\'omica, Molecular y Nuclear,
Universidad de Granada, Granada 18071, Spain}

\affiliation{$^b$Dipartimento di Fisica Teorica, Universit\`a di Torino and
  INFN, Sezione di Torino, Via P. Giuria 1, 10125 Torino, Italy}

\affiliation{$^c$Departamento de F\'{\i}sica At\'omica, Molecular y Nuclear,
Universidad de Sevilla, Apdo.1065, 41080 Sevilla, Spain}

\affiliation{$^d$Center for Theoretical Physics, Laboratory for Nuclear
  Science and Department of Physics, Massachusetts Institute of Technology,
  Cambridge, MA 02139, USA}

\date{\today}


\begin{abstract}
Two-particle two-hole contributions to electromagnetic response
functions are computed in a fully relativistic Fermi gas model. All
one-pion exchange diagrams that contribute to the scattering
amplitude in perturbation theory are considered, including terms for
pionic correlations and meson-exchange currents (MEC).  The pionic
correlation terms diverge in an infinite system and thus are
regularized by modification of the nucleon propagator in the medium
to take into account the finite size of the nucleus.  The pionic
correlation contributions are found to be of the same order of
magnitude as the MEC.

\end{abstract}


\maketitle

\section{Introduction}

The goal of this paper is to present a fully relativistic
calculation of the two-particle two-hole (2p-2h) contributions to
the inclusive $(e,e')$ response functions of nuclei for intermediate
to high momentum transfers in a Fermi gas model. Consistency with
perturbation theory is maintained and all diagrams with one-pion
exchange in the nuclear current are considered, constructed by
attaching a photon to all possible lines in the basic one-pion
exchange Feynman diagram. In this way not only meson-exchange
currents (MEC) arise (for example, where the photon is attached to
the pion), but also pionic correlation diagrams, where the virtual
photon is absorbed by one of the two interacting nucleons. Both
kinds of diagrams are considered in our model, together with the
usual virtual $\Delta$-isobar electroexcitation and decay.

We are motivated by previous work presented
in~\cite{DePace03,DePace04}, where only the MEC were included in the
2p-2h transverse ($T$) response, together with earlier work both in
non-relativistic~\cite{Van80} and
relativistic~\cite{Dek91,Dek92,Dek94,Dek95} regimes.  The
contribution found from the 2p-2h excitations is small at the
quasielastic (QE) peak, and increases with energy transfer, being
more important in the dip region, where it is dominated by the
$\Delta$ current.  At the non-relativistic level attempts were also
made to evaluate the 2p-2h contribution of MEC in the $T$-response
for finite nuclei in a shell model~\cite{Ama93,Ama94}.

The MEC are not the only two-body operators able to induce 2p-2h
excitations. The correlation operators arising from Feynman diagrams
where the photon is attached to a nucleon line, exchanging a pion
with another nucleon, are of the same order as the MEC in the
perturbative expansion and should be included to be
consistent~\cite{Alb84,Alb91,Gil97}.  These diagrams, however,
present the problem of giving an infinite answer in a Fermi gas
model. The reason is that there is a nucleon propagator that can be
on-shell in the region of the quasielastic peak. Since the response
function is the square of the amplitude, the resulting double pole
gives an infinite result after integration. In dealing with this
problem, in~\cite{Alb84} a prescription was followed by keeping the
lines with a nucleon propagator strictly off the mass shell. A
different approach was taken in~\cite{Alb91} by subtracting from the
proper self-energy its value on the mass shell, with the unphysical
shortcoming of obtaining negative results for the 2p-2h responses to
the left of the QE peak. Finally, in \cite{Gil97} a nucleon
self-energy in the medium was introduced in the nucleon propagator.
In dealing with the seven-dimensional integrals appearing in the
2p-2h responses, some of the previous calculations have resorted to
the approximation of setting the two hole momenta both equal to zero
in some of the diagrams~\cite{Alb84} or by taking into account only
an average nucleon momentum~\cite{Gil97}.

In this work we revisit the double-pole problem to analyze the
nature of the divergence of the resulting contributions. By
isolating the divergent terms we are able to link them to the
infinite extension of the Fermi gas system. In fact the double-pole
term can be related to the probability of one-nucleon emission
followed by nucleon re-scattering off another nucleon, with the
final ejection of two particles. This probability is infinite, since
it is proportional to the propagation time of a real nucleon in a
Fermi gas.  This fact was pointed out in \cite{Gil97} where it was
cured, as mentioned above, by introducing a nucleon self-energy with
an imaginary part giving it a finite lifetime for collisions. In
this paper we use a similar procedure by introducing a finite
imaginary part $i\epsilon$ in the nucleon propagator, but with a new
meaning for the free parameter $\epsilon$. Instead of being an
imaginary part of the nucleon self-energy for collisions, we relate
it to the time $T$ that a nucleon can travel across the nucleus
before leaving it. Hence this term accounts for the finite size of a
real nucleus in contrast to an infinite system like the Fermi gas,
where $T$ is infinite. The value of $\epsilon$ can be estimated to
be roughly about 200 MeV, appreciably larger than the usual values
of the nucleon width for collisions.

The structure of this work is as follows. In Sect.~II we present our
model and define the 2p-2h response functions and the two-body
current operators. We discuss in depth the divergence of the
correlation diagrams and the need to introduce the parameter
$\epsilon$ in Sect.~III (details of the numerical calculation are
given in the appendices). In Sect.~IV we present results for the
2p-2h longitudinal and transverse response functions. In the case of
the correlation diagrams we present results for several values of
the parameter $\epsilon$. Finally, in Sect.~V we present our
conclusions.

\section{Model for 2p-2h response functions}

We consider an electron that scatters off a nucleus transferring
four-momentum $Q^{\mu}=(\omega,\nq)$, with $\omega$ the energy
transfer and $\nq$ the momentum transfer. We follow closely the
notation of \cite{Ama02}. Assuming plane waves for the electron,
working in the laboratory system and taking the $z$ direction along
the momentum transfer, the inclusive cross section is written as
\begin{equation}
\frac{d\sigma}{d\Omega'_e d\omega}
=\sigma_M
\left[
v_LR_L(q,\omega)
+v_T R_T(q,\omega)
\right] \, ,
\end{equation}
where $\sigma_M$ is the Mott cross section, $v_L$ and $v_T$ are the
lepton kinematic factors, and the relevant quantities are the
longitudinal $R_L(q,\omega)$ and transverse $R_T(q,\omega)$ response
functions, respectively.  These are defined as the following
components of the hadronic tensor,
\begin{eqnarray}
R_L&=& W^{00}\\
R_T&=& W^{11}+W^{22} \, ,
\end{eqnarray}
where
\begin{equation}
W^{\mu\nu}=\sum_f
\langle f |J^{\mu}(Q)|i\rangle^*
\langle f |J^{\nu}(Q)|i\rangle
\delta(E_i+\omega-E_f)
\end{equation}
and $J^{\mu}(Q)$ is the nuclear current operator.

In this paper we take the initial nuclear state as the relativistic
Fermi gas (RFG) model ground state, $|i\rangle=|F\rangle$, with all
states with momenta below the Fermi momentum $k_F$ occupied. The sum
over final states can be decomposed as the sum of one-particle
one-hole (1p-1h) plus two-particle two-hole (2p-2h) excitations plus
additional channels. In the impulse approximation the 1p-1h channel
gives the well-known response functions of the RFG. Here we focus on
the 2p-2h channel where the final states are of the type
$|f\rangle=|\np'_1 s'_1, \np'_2 s'_2, \nh_1^{-1} s_1,  \nh_2^{-1}
s_2 \rangle$, where $\np'_i$ are momenta of relativistic final
nucleons above the Fermi sea, $p'_i>k_F$, with four-momenta
$P'_i=(E'_i,\np'_i)$, and $H_i=(E_i,\nh_i)$ are the four-momenta of
the hole states with $h_i<k_F$. The spin indices are $s'_i$ and
$s_i$.

\subsection{2p-2h Response functions}

Since we have two species of nucleons, the 2p-2h responses can be
further decomposed as the sum of two-proton (PP),
two-neutron (NN) and proton-neutron (PN) emission
\begin{equation}
R_K=R_K(PP)+R_K(NN)+R_K(PN) \, .
\end{equation}
For the PP channel we write down
the L response as (likewise for the T response):
\begin{eqnarray}
R_L(PP)&=&
\nonumber \\
&&\kern -1cm
\frac14
\sum_{\np'_1s'_1}
\sum_{\np'_2s'_2}
\sum_{\nh_1s_1}
\sum_{\nh_2s_2}
\left| \langle
\np'_1\np'_2\nh_1^{-1}\nh_2^{-1}|J^0(Q)|F\rangle
\right|^2
\nonumber \\
&&
\kern -1cm \mbox{}
\times\delta(E'_1+E'_2-\omega-E_1-E_2)
\, ,
\end{eqnarray}
where the spin indices are implicit in the matrix elements. The
factor $\frac14$ comes from anti-symmetry of the wave functions, to
avoid double counting of the final states under the interchange of
the indices $1'\leftrightarrow 2'$ and $1\leftrightarrow 2$.
Exploiting the anti-symmetry, the many-body matrix element of a
two-body operator can be written as the direct minus exchange part
of the two-body current matrix element
\[
\langle\np'_1\np'_2\nh_1^{-1}\nh_2^{-1}|J^\mu|F\rangle=
\langle\np'_1\np'_2 |J^\mu|\nh_1\nh_2\rangle -\langle\np'_1\np'_2
|J^\mu|\nh_2\nh_1\rangle \, ,
\]
which we write  in terms of the two-body current function
$j^{\mu}(\np'_1,\np'_2,\nh_1,\nh_2)$ to be specified below,
\begin{eqnarray}
&&\langle\np'_1\np'_2 |J^\mu|\nh_1\nh_2\rangle=
(2\pi)^3\delta(\np'_1+\np'_2-\nh_1-\nh_2-\nq)
\nonumber\\
&&\times\frac{m^2}{V^2(E_1E_2E'_1E'_2)^{1/2}}
j^{\mu}(\np'_1,\np'_2,\nh_1,\nh_2).
\end{eqnarray}
Going to the thermodynamic limit and integrating over the momentum $\np'_2$
we obtain
\begin{eqnarray}
&&R_L(PP)=
\frac{V}{4}
\sum_{s'_1s'_2s_1s_2}
\int
\frac{d^3p'_1}{(2\pi)^3}
\frac{d^3h_1}{(2\pi)^3}
\frac{d^3h_2}{(2\pi)^3}
\nonumber\\
&&\times
\frac{m^4}{E_1E_2E'_1E'_2}
\left|j^{0}(\np'_1,\np'_2,\nh_1,\nh_2)_A\right|^2
\nonumber\\
&&\mbox{}\times\delta(E'_1+E'_2-\omega-E_1-E_2)
\theta(p'_2-k_F)\, ,
\label{integral}
\end{eqnarray}
where $\np'_2=\nh_1+\nh_2+\nq-\np'_1$, and
the integration limits  are $h_1,h_2<k_F$, $p'_1>k_F$.
We have defined the anti-symmetrized current function
\[
j^{\mu}(1',2',1,2)_A\equiv
j^{\mu}(1',2',1,2)-
j^{\mu}(1',2',2,1)
\]
\begin{figure}[tph]
\begin{center}
\includegraphics[scale=0.65,  bb= 130 470 490 690]{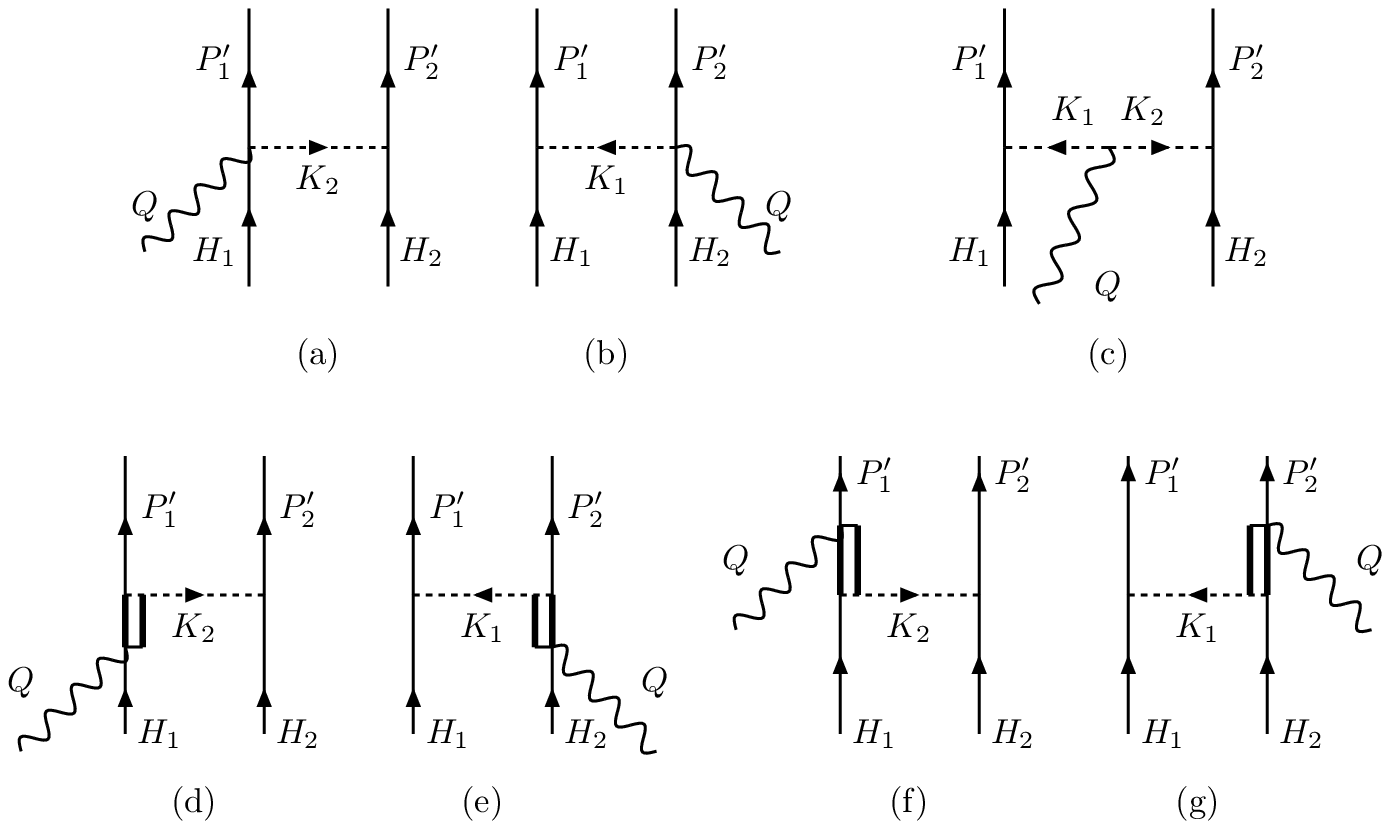}
\caption{
MEC diagrams considered in the present study.
Diagrams (a,b) correspond to the seagull, (c) to the pionic, and (d-g)
to the $\Delta$ current, respectively.
}
\label{Fig1}
\end{center}
\end{figure}
with obvious meaning for the abbreviated arguments. Expanding the
square inside the integral in Eq.~(\ref{integral}), three terms are
obtained:
\begin{eqnarray}
\left|j^{\mu}(1',2',1,2)_A\right|^2&=&
\left|j^{\mu}(1',2',1,2)\right|^2
+\left|j^{\mu}(1',2',2,1)\right|^2
\nonumber\\
&-&
2{\rm Re}\ j^{\mu}(1',2',2,1)^*j^{\mu}(1',2',1,2)\, .
\nonumber\\
\end{eqnarray}
Changing variables $1\leftrightarrow 2$ in the second term under the
integral, we obtain the first term again. Hence we can finally write
for the PP response
\begin{eqnarray}
R_L(PP)&=&
\frac{V}{2}
\sum_{s'_1s'_2s_1s_2}
\int
\frac{d^3p'_1}{(2\pi)^3}
\frac{d^3h_1}{(2\pi)^3}
\frac{d^3h_2}{(2\pi)^3}
\nonumber\\
&&
\kern -1cm
\times
\frac{m^4}{E_1E_2E'_1E'_2}
\left[
\left|j^{0}(\np'_1,\np'_2,\nh_1,\nh_2)\right|^2
\right.
\nonumber\\
&&
\kern -1cm
\left.
-{\rm Re}\
j^{0}(\np'_1,\np'_2,\nh_1,\nh_2)^*
j^{0}(\np'_1,\np'_2,\nh_2,\nh_1)
\right]
\nonumber\\
&&
\kern -1cm
\mbox{}\times\delta(E'_1+E'_2-\omega-E_1-E_2)
\theta(p'_2-k_F) \,.
\label{RL}
\end{eqnarray}
Note that the factor $\frac12$ in front of the sum comes from the
anti-symmetry of the particles (protons).  A similar expression is
obtained for the NN response $R_L(NN)$. In the case of the PN
channel we subtract the charge exchange contribution without any
symmetry term because there are no additional isospin sums, and the
result is
\begin{eqnarray}
&&R_L(PN)=
V
\sum_{s'_1s'_2s_1s_2}
\int
\frac{d^3p'_1}{(2\pi)^3}
\frac{d^3h_1}{(2\pi)^3}
\frac{d^3h_2}{(2\pi)^3}
\nonumber\\
&&\times
\frac{m^4}{E_1E_2E'_1E'_2}
\left|
\langle PN|j^{0}(\np'_1,\np'_2,\nh_1,\nh_2)|PN\rangle
\right.
\nonumber\\
&&
\left.
-\langle NP|j^{0}(\np'_1,\np'_2,\nh_2,\nh_1)|PN\rangle
\right|^2
\nonumber\\
&&\mbox{}\times\delta(E'_1+E'_2-\omega-E_1-E_2)
\theta(p'_2-k_F) \, .
\label{RLpn}
\end{eqnarray}
Finally, note that the 2p-2h response is proportional to the volume
of the system V which is related to the number of particles ${\cal
N}$ (protons or neutrons) by $V=3\pi^2 {\cal N}/k_F^3$.

\subsection{Two-body current matrix elements}

\begin{figure}[tph]
\begin{center}
\includegraphics[scale=0.65,  bb= 130 580 490 690]{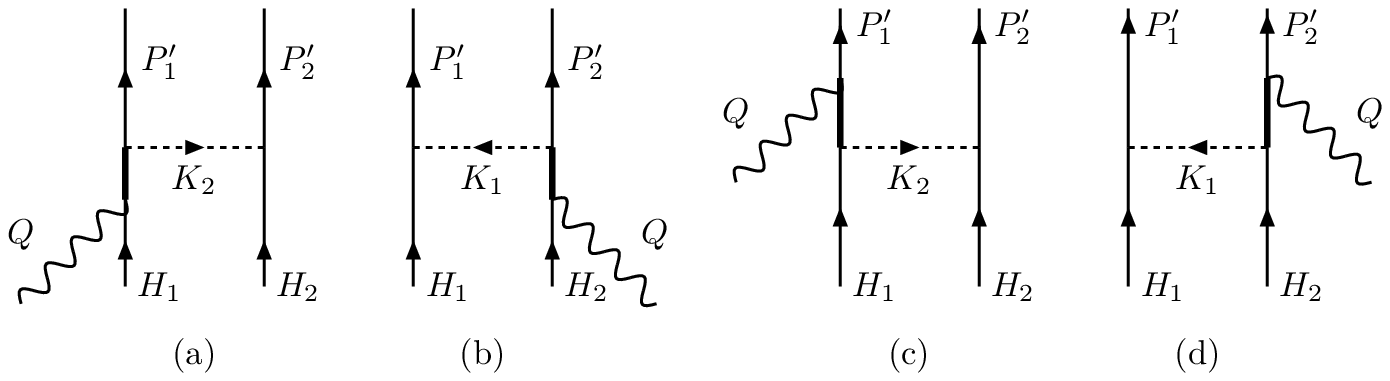}
\caption{
Correlation diagrams considered in the present study.
Diagrams (a,b) correspond to the forward, and (c-d)
backward contributions, respectively.
}
\label{Fig2}
\end{center}
\end{figure}

The MEC considered in this work are represented by the Feynman
diagrams of Fig.~1. The pionic four-momenta $K_1$, $K_2$ are defined
via $K^\mu_i=P'_i{}^\mu-H^\mu_i$ as the four-momenta given to the
nucleons 1 and 2, respectively, by the exchanged pion.

Assuming pseudo-vector nucleon-pion coupling, the fully relativistic
two-body current matrix elements are given by~\cite{Ama02,Ama03}:

\begin{itemize}
\item
(a-b) Seagull or contact:
\end{itemize}
\begin{eqnarray}
j^{\mu}_{s}(\np'_1, \np'_2,\np_1,\np_2)
&=& \frac{f^2}{m_\pi^2}
             i\epsilon_{3ab}
             \overline{u}(\np'_1)\tau_a\gamma_5\Kbar_1 u(\np_1)
\nonumber\\
&&\kern -3cm \times
             \frac{F_1^V}{K_1^2-m_\pi^2}
             \overline{u}(\np'_2)\tau_b\gamma_5\gamma^{\mu}u(\np_2)
             + (1 \leftrightarrow 2) \,.
\label{s}
\end{eqnarray}

\begin{itemize}
\item
(c) Pion-in-flight:
\end{itemize}
\begin{eqnarray}
j^{\mu}_{p}(\np'_1, \np'_2,\np_1,\np_2)
&=&
\frac{f^2}{m_\pi^2}
             i\epsilon_{3ab}
             \frac{F_\pi(K_1-K_2)^\mu}{(K_1^2-m_\pi^2)(K_2^2-m_\pi^2)}
\nonumber\\
&&\kern -3cm \times
     \overline{u}(\np'_1)\tau_a\gamma_5\Kbar_1 u(\np_1)
             \overline{u}(\np'_2)\tau_b\gamma_5\Kbar_2 u(\np_2) \, .
\label{p}
\end{eqnarray}
In the above we use the Einstein convention for the sum over a
repeated isospin index $a=1,2,3$. Moreover, $F_1^V$ and $F_\pi$ are
the electromagnetic isovector nucleon and pion form factors,
respectively.  The spinors are normalized according to the Bjorken
and Drell convention~\cite{Bjo65} and the pion-nucleon coupling
constant is $f^2/4\pi = 0.08$.

\begin{itemize}
\item (d-g) Delta current:
\end{itemize}
\begin{eqnarray}
j^{\mu}_{\Delta}(\np'_1, \np'_2,\np_1,\np_2)
&=&
\frac{f_{\pi N\Delta} f}{m_\pi^2}
\frac{1}{K_2^2-m_\pi^2}
             \overline{u}(\np'_1)T_a^\mu(1) u(\np_1)
\nonumber\\
&&\kern -2cm
\times
             \overline{u}(\np'_2)\tau_a\gamma_5\Kbar_2 u(\np_2)
             + (1 \leftrightarrow 2) \, .
\label{d}
\end{eqnarray}
The vector $T_a^{\mu}(1)$ is related to the pion electroproduction
amplitude
\begin{eqnarray}
T_a^\mu(1)
&=&
K_{2,\alpha}
\Theta^{\alpha\beta}
G^\Delta_{\beta\rho}(H_1+Q)
S_f^{\rho \mu}(H_1)
T_a T_3^{\dagger}
\nonumber\\
&&+
T_3 T_a^{\dagger}
S_b^{\mu \rho}(P'_1)
G^\Delta_{\rho\beta}(P'_1-Q)
\Theta^{\beta\alpha}
K_{2,\alpha} \, .
\end{eqnarray}
The forward $\Delta$ electroexcitation
tensor is~\footnote{Note that there is a sign error in Eq.~(15) of
\cite{Ama03}.}
\begin{eqnarray}
S_f^{\rho \mu}(H_1)
&=&
\Theta^{\rho\mu}     \left[
       g_1  \Qbar
     - g_2 H_1\cdot Q
      +g_3Q^2
\right]
\gamma_5
\nonumber\\
&-&
\Theta^{\rho\nu}Q_\nu
\left[
       g_1 \gamma^\mu
      -g_2 H_1^\mu
      + g_3Q^\mu
\right]
\gamma_5
\end{eqnarray}
and the backward tensor amplitude is
\begin{eqnarray}
S_b^{\rho \mu}(P'_1)
&=&
\gamma_5
\left[
       g_1  \Qbar
     -g_2P'_1\cdot Q
     -g_3Q^2
\right]
\Theta^{\mu\rho}
\nonumber\\
&-&
\gamma_5
\left[
       g_1\gamma^\mu
      -g_2P'_1{}^{\mu}
      -g_3 Q^\mu
\right]
Q_\nu\Theta^{\nu\rho} \, .
\end{eqnarray}
The tensor $\Theta_{\mu\nu}$ is defined by
\begin{equation}
\Theta_{\mu\nu}=g_{\mu\nu}-\frac14\gamma_\mu \gamma_\nu \, .
\label{Theta}
\end{equation}
For the $\Delta$ propagator
we use  the usual Rarita-Schwinger (RS) tensor
\begin{eqnarray}
 G^{\Delta}_{\beta\rho}(P)
&=&
 -\frac{ \Pbar+m_\Delta}{P^2-m_\Delta^2}
\nonumber\\
&&
\kern -3cm
\times
\left[
         g_{\beta\rho}
        - \frac{1}{3} \gamma_\beta\gamma_\rho
        - \frac{2}{3} \frac{P_\beta P_\rho}{m_\Delta^2}
        - \frac{\gamma_\beta P_\rho - \gamma_\rho P_\beta}{3m_\Delta}
\right] \, .
\end{eqnarray}
In what follows we perform the substitution $m_\Delta\rightarrow
m_\Delta+\frac{i}{2}\Gamma(P)$ in the denominator of the propagator
to account for the $\Delta$ decay probability. Finally, the
electromagnetic coupling constants $g_i$ are given by
\begin{equation}
g_1=\frac{G_1}{2 m_N}\, ,
\kern 1cm
g_2=\frac{G_2}{4 m_N^2} \, ,
\kern 1cm
g_3=\frac{G_3}{4 m_N^2} \, .
\end{equation}
Our approach for the $\Delta$ follows, as a particular case, from
the more general form of the $\gamma N \Delta$ Lagrangian of
Pascalutsa {\it et al.}~\cite{Pas95}. The $\Delta$ coupling
constants used here are $G_1=4.2$, $G_2=4$, $G_3=1$, and $f_{\pi
N\Delta} = 4\times 0.564$.

The correlation current is defined in Fig.~2, and given by
\begin{eqnarray}
j^{\mu}_{cor}(\np'_1, \np'_2,\np_1,\np_2)
&=&           \frac{f^2}{m_\pi^2}
              \overline{u}(\np'_1)\tau_a\gamma_5\Kbar_1 u(\np_1)
              \frac{1}{K_1^2-m_\pi^2} \nonumber\\
& & \kern -2cm
    \mbox{}\times \overline{u}(\np'_2)
    \left[    \tau_a\gamma_5\Kbar_1
              S_F(P_2+Q)\Gamma^\mu(Q)
     \right.
\nonumber\\
&&\kern -1cm
    \left.
            + \Gamma^\mu(Q)S_F(P'_2-Q)
              \tau_a\gamma_5\Kbar_1
    \right]u(\np_2) \nonumber\\
& & \kern -2cm \mbox{}+ (1\leftrightarrow2)\, ,
\label{correlation}
\end{eqnarray}
where $S_F(P)$ is the Feynman propagator for the nucleon
\begin{equation}
S_F(P) =\frac{\Pbar + m}{P^2-m^2+i\epsilon}
\end{equation}
and $\Gamma^\mu(Q)$ is the electromagnetic nucleon vertex,
\begin{equation}
\Gamma^\mu(Q) =
F_1\gamma^\mu+\frac{i}{2m}F_2\sigma^{\mu\nu}Q_\nu \, .
\end{equation}
The nucleon form factors $F_1$ and $F_2$
are given by the Galster parametrization~\cite{Gal71}.

The isospin sums and isospin matrix elements must be performed
separately for each isospin channel. Explicit expressions are given
in Appendix~A.

\section{Divergence of the correlation responses}

The response functions computed using the correlation current in
Eq.~(\ref{correlation}) are divergent in the Fermi gas. There are
two sources for this divergence: the first one comes from the double
pole of the propagator when taking the square of the current. This
divergence can be shown to behave as $1/\epsilon$ plus principal
value terms going as $\log\epsilon$. The second source is related to
the behavior of the principal values arising from the double and
single poles near the RFG boundary of the quasielastic peak, where
the principal values present a logarithmic divergence.

To illustrate the mathematical structure of this divergence
we isolate as an example the singularities produced by the
diagram of Fig.~2(a). The
corresponding current operator can be written as
\begin{equation}
j^\mu=\frac{l^{\mu}}{E_1+\omega-E_{\nh_1+\nq}+i\epsilon} \, ,
\end{equation}
where
$E_{\np}=\sqrt{m^2+\np^2}$ is the on-shell energy.
We have explicitly extracted the divergent part of the denominator,
with  a pole for
\begin{equation}
E_{\nh_1+\nq}=E_1+\omega
\label{condition}
\end{equation}
in the limit $\epsilon\rightarrow 0$.
The above equation is equivalent to the quasielastic condition for
emission of an on-shell nucleon with four-momentum $H_1+Q$.
In fact, for a given value of $h_1$, Eq.~(\ref{condition}) holds
when the angle between $\nh_1$ and $\nq$ is given by
\begin{equation}\label{quasielastic}
\cos\theta_1=\frac{Q^2+2E_1\omega}{2h_1q} \, .
\end{equation}
Since the condition  $-1<\cos\theta_1< 1$ defines the
boundary of the quasielastic peak, the pole can always
be reached in that region.

To study the behavior of the response functions due to this pole, it
is convenient to change the variable $\theta_1$ to a new variable
defined by
\begin{eqnarray}
x_1\equiv E_1+\omega-E_{\nh_1+\nq}
\end{eqnarray}
in the integral over $\nh_1$ in Eq.~(\ref{RL}). Then the components
of the total current matrix element can be written as a function of
$x_1$ in the general form
\begin{equation}
f(x_1)=\frac{\varphi(x_1)}{x_1+i\epsilon}+g(x_1) \, ,
\end{equation}
where the first term comes from diagram 2(a) and the function
$g(x_1)$ comes from the sum of the remaining diagrams, and is finite
for $x_1=0$. Since the current appears squared in the response
function, we are dealing with the integral of a function of the kind
\begin{equation}
|f(x_1)|^2=\frac{|\varphi(x_1)|^2}{x_1^2+\epsilon^2}+|g(x_1)|^2
+2 {\rm Re}\ \frac{\varphi^*(x_1)g(x_1)}{x_1-i\epsilon} \, .
\label{square}
\end{equation}
When integrating this function over $x_1$, and taking the limit
$\epsilon\rightarrow 0$, the first term has a double pole for $x_1=0$,
while the third one has a single pole. To deal with the single pole
we use the usual Plemeli relation,
\begin{equation}
\frac{1}{x+i\epsilon}= {\cal P}\ \frac{1}{x}-i\pi\delta(x) \, .
\label{dirac}
\end{equation}
To apply a similar relation for the double pole term, we add and
subtract the on-shell value $|\varphi(0)|^2/(x_1^2+\epsilon^2)$.
Taking the limit $\epsilon\rightarrow 0$ we can use relations which
are valid for any function $\psi(x)$
\begin{equation}
\int^b_{-a}
\frac{\psi(x)-\psi(0)}{x^2+\epsilon^2}dx
\rightarrow {\cal P}\
\int^b_{-a}
\frac{\psi(x)- \psi(0)}{x^2}dx
\end{equation}
and
\begin{equation}
\int^b_{-a} \frac{\psi(0)}{x^2+\epsilon^2}dx = \frac{1}{\epsilon}
\left[\tan^{-1}\frac{b}{\epsilon}+\tan^{-1}\frac{a}{\epsilon}
\right] \psi(0) \sim \frac{\pi}{\epsilon}\psi(0) \, .
\end{equation}
Then Eq.~(\ref{square}) can be written in the form
\begin{eqnarray}
|f(x_1)|^2
&=&
{\cal P}\ \frac{|\varphi(x_1)|^2-|\varphi(0)|^2}{x_1^2}
+|g(x_1)|^2
\nonumber\\
&+&2 {\cal P}\
\frac{{\rm Re}\ \varphi^*(x_1)g(x_1)}{x_1}
-2\pi{\rm Im}\ \varphi^*(0)g(0)\delta(x_1)
\nonumber\\
&+&
\frac{|\varphi(0)|^2}{\epsilon}\pi\delta(x_1) \, .
\label{expansion}
\end{eqnarray}
The last $O(1/\epsilon)$ term in Eq.~(\ref{expansion}) provides the
dominant contribution to the response function, being infinite for
$\epsilon\rightarrow 0$. Due to the $\delta$ function, that term
does not contribute outside the quasielastic-peak region, where
$x_1$ is different from zero.

The principal values present in Eq.~(\ref{expansion}) also diverge
in the particular case in which one of the limits of integration is
zero. In that case the principal value in Eq.~(\ref{dirac}) should
be computed instead using
\begin{equation}
{\cal P}\int_{-a}^b\frac{\psi(x)}{x}dx=
\int_{-a}^b\frac{\psi(x)-\psi(0)}{x}dx
+\frac12\psi(0)\ln\frac{b^2+\epsilon^2}{a^2+\epsilon^2}
\end{equation}
and it gives a $\ln\epsilon$ term if $a$ or $b$ is zero.  That
situation in fact occurs throughout the quasielastic region, and in
particular at the boundary of the quasielastic peak.  Therefore one
expects an additional divergence $\sim O(\ln\epsilon)$.

The meaning of the term
$\frac{|\varphi(0)|^2}{\epsilon}\pi\delta(x_1)$ is explained in what
follows.  Diagram 2(a), when the intermediate nucleon is on shell,
gives the probability of a 1p-1h electroexcitation times the
probability of quasielastic nucleon scattering. Since the
interaction probability is proportional to the interaction time $T$,
the probability of this re-scattering process is proportional to
$T^2$. Therefore the cross section is proportional to $T$.  In an
infinite system such as the Fermi gas, the intermediate nucleon
never leaves the nucleus and therefore $T\rightarrow \infty$.
However, in a finite nucleus one expects no divergence because a
high-energy nucleon will leave the nucleus in a finite time.
Therefore the interaction time is finite.

The relation between $\epsilon$ and $T$ can also be obtained by
inspection of the momentum-space propagator in quantum field
theory~\cite{Man84}, computed as the vacuum expectation value of
time-ordered Fermion fields. The value $\epsilon$ in the denominator
of the propagator can be seen as a regularization parameter in the
Fourier transform of the time step function for a particle with
four-momentum $P^\mu=(p_0,\np)$
\begin{equation}\label{propagator}
\int_{-T/2}^{T/2} dt \,
{\rm  e}^{i(p_0-E_{\np})t}\theta(t)
= \frac{i}{p_0-E_{\np}+i\epsilon} \, ,
\end{equation}
where $T\rightarrow\infty$ and $\epsilon\rightarrow0$.
For a real particle, $p_0-E_{\np}=0$, the left-hand side of
the above equation is $T/2$, and the right-hand side is $1/\epsilon$.
Therefore
\begin{equation}
\frac{T}{2}= \frac 1\epsilon \, .
\end{equation}
This can be obtained alternatively by replacing
 the on-shell value of the propagator in Eq.~(\ref{propagator})
as a delta function
\begin{equation}
\frac{1}{\epsilon}=
\lim_{p_0\rightarrow E_{\np}}
\frac{i}{p_0-E_{\np}+i\epsilon}
=
\pi\delta(0)
\end{equation}
and using the integral representation
\begin{equation}
\delta(0)
= \lim_{T\rightarrow\infty}
\frac{1}{2\pi}\int_{-T/2}^{T/2}dt= \frac{T}{2\pi}.
\end{equation}
In this paper we cure the divergence of the correlation diagram by a
regularization procedure, using a finite value for $\epsilon$ to
account for the finite propagation time of a high-energy nucleon in
a nucleus before leaving it. To estimate the value of $\epsilon$ for
a nucleus such as $^{12}$C, we assume that the nucleon moves at the
velocity of light and it has to cross a distance equal to the
nuclear radius $R\sim 2$~fm. Then
\begin{equation}\label{epsilon}
\epsilon\simeq \frac{2\hbar}{T} \simeq \frac{2\hbar c}{R}\simeq
\frac{400}{2}{\rm MeV}\simeq 200\, {\rm MeV} \, .
\end{equation}
Note that this value, $\epsilon\simeq 200$ MeV, is very different
from the nucleon width $\Gamma \sim 10$ MeV which is usually
obtained in nuclear matter as the width for nuclear inelastic
interaction. In practice the value of $\epsilon$ can be taken as a
parameter to be fitted to data. In the next section we perform a
study of the dependence of our results upon $\epsilon$. Unless
otherwise specified we assume $\epsilon=200$ MeV.

At this point we should mention that the use of Eq.~(\ref{square})
to compute the 2p-2h response functions becomes impractical due to
complications in the numerical calculation of principal values in
multidimensional integrals including the four diagrams of Fig. 2
(and the corresponding exchange parts). Since we are forced to use a
finite value of $\epsilon$, it becomes more convenient to keep from
the beginning the $i\epsilon$ term in the denominator of the nucleon
propagator in Eq.~(\ref{correlation}).

\begin{figure}[tph]
\begin{center}
\includegraphics[scale=0.75,  bb= 220 275 400 770]{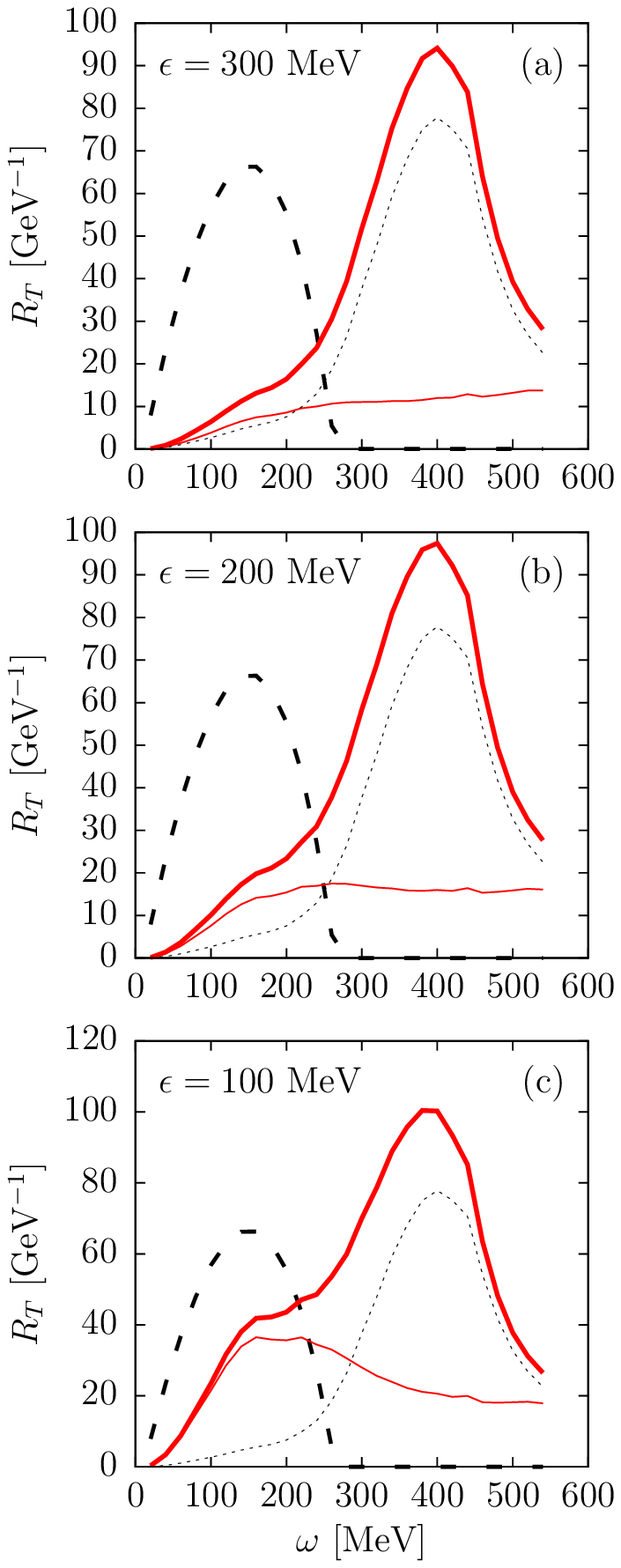}
\caption{ 2p-2h transverse response of $^{56}$Fe at $q=550$ MeV/c.
Three values
  of the parameter $\epsilon$ are shown.  Thin solid lines:
  Correlation only. Dotted lines: MEC only. Thick solid lines: total.
  Dashed: RFG OB results.  }
\label{Fig3}
\end{center}
\end{figure}

\begin{figure}[th]
\begin{center}
\includegraphics[scale=0.75,  bb= 220 275 400 770]{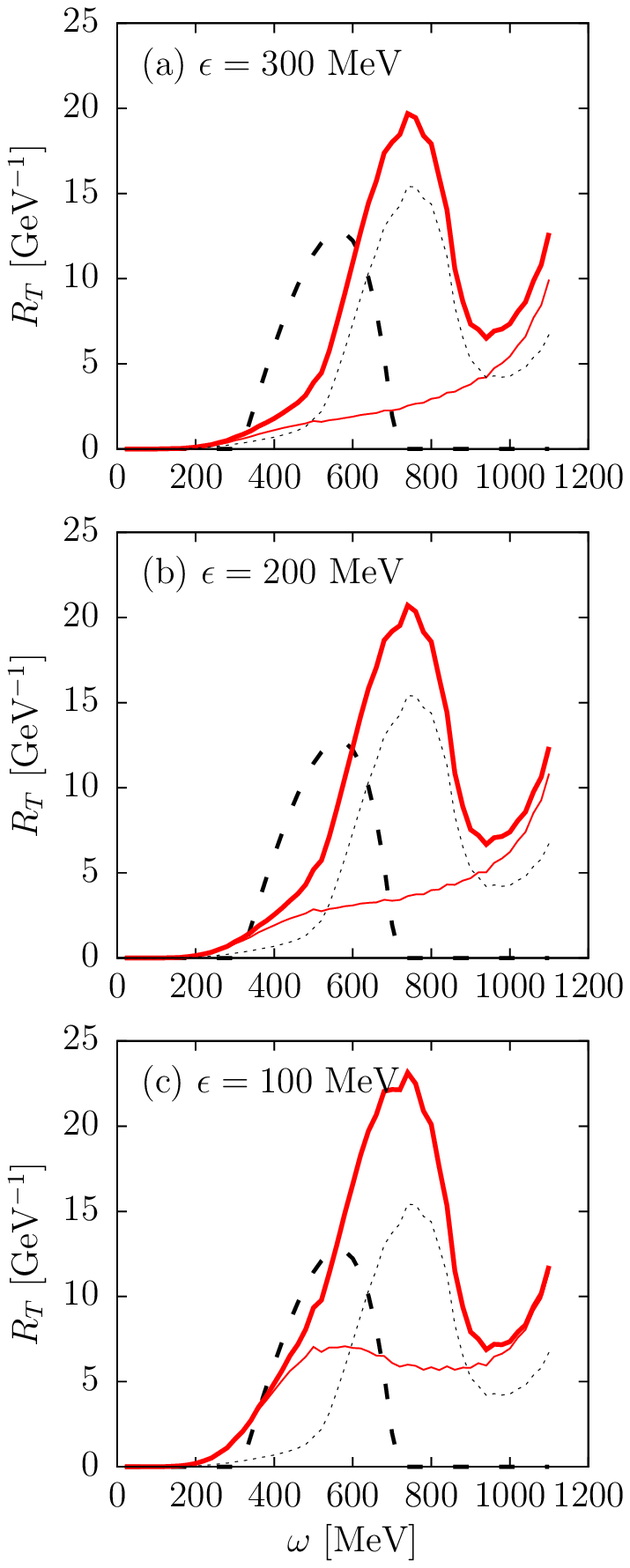}
\caption{ As for Fig. 3, but now at $q=1140$ MeV/c.} \label{Fig4}
\end{center}
\end{figure}

\section{Results}

Here we present results for the longitudinal and transverse response
functions for inclusive two-particle emission.  We compute the 2p-2h
response functions in the RFG model as the 9-dimensional integrals
given by Eqs.~(\ref{RL},\ref{RLpn}).  The energy delta function can
be used to integrate over $p'_1$, fixing the energy $E'_1$ of the
first particle. More details are given in Appendix~B. By rotational
invariance considerations, one of the the azimuthal angles can be
fixed, multiplying at the same time the result by a factor $2\pi$.
We choose $\phi'_1=0$. At the end we have a 7-dimensional
integration to be performed numerically. The usual procedure is to
use a multi-dimensional Montecarlo (MC) integration. Since the pole
structure of the integrand is numerically delicate, in this work we
use instead a mixed Montecarlo-Simpson integration procedure. The
Simpson algorithm is used for integration over the angles of the two
holes $\theta_1,\theta_2$ and of the first particle $\theta'_1$. The
remaining 4-dimensional integral over the hole momenta $h_1,h_2$ and
their angles $\phi_1,\phi_2$ is made by Montecarlo. To keep the CPU
times manageable we use a number of MC points of the order of $10^3$
for $q=1$ GeV/c.  For other values of the momentum transfer the
number of MC points is modified linearly with $q$. We have performed
a study of the stability of the results with the number of MC points
and have found that the error from the integration procedure is
within a few percent.

A pion-nucleon form factor is included in the 2-body currents:
$F_{\pi NN}(K_\pi)=(\Lambda^2-m_\pi^2)/(\Lambda^2-K_\pi^2)$, with
$\Lambda=1.3$ GeV. We use the same value for the $\pi N\Delta$ form
factor in the Delta current. The electromagnetic form factors are
those of Galster for the nucleon, and those used
in~\cite{Ama02,Ama03} for the MEC.

\begin{figure}[th]
\begin{center}
\includegraphics[scale=0.75,  bb= 220 275 400 770]{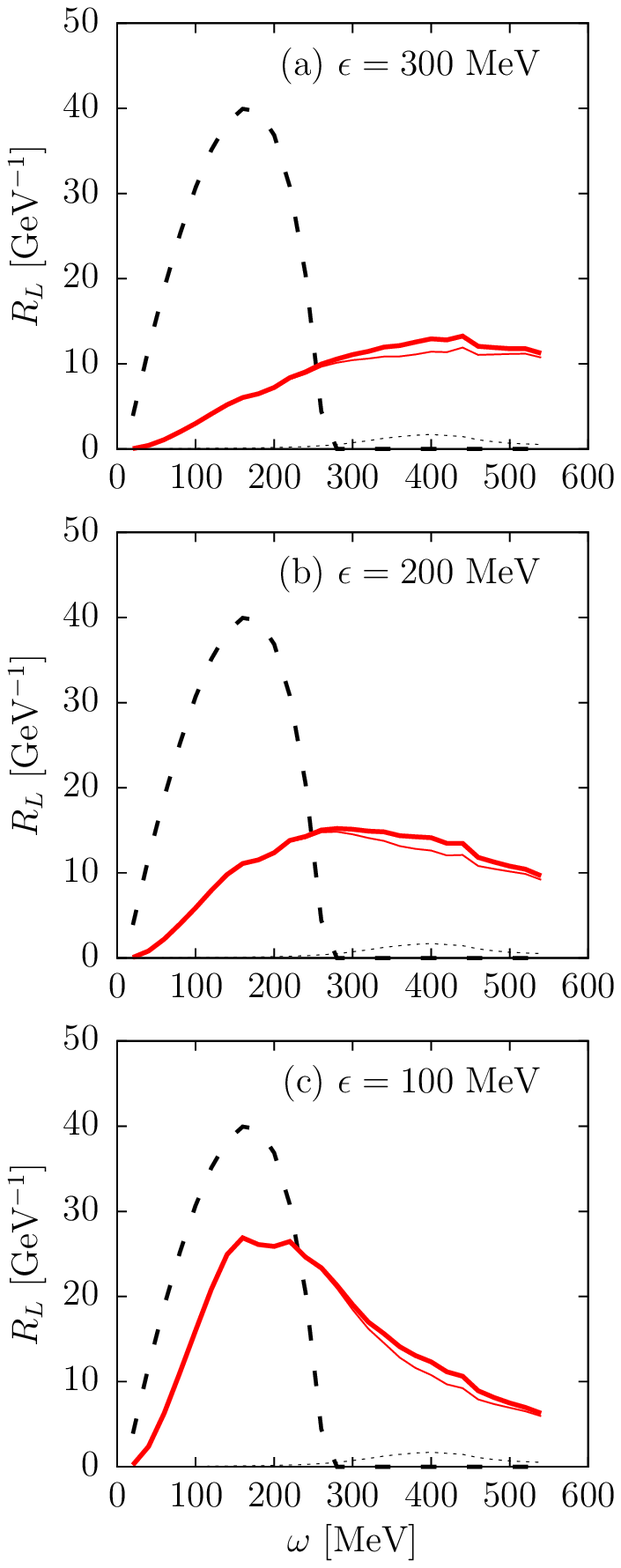}
\caption{ As for Fig. 3, but now for $R_L$.} \label{Fig5}
\end{center}
\end{figure}

\begin{figure}[th]
\begin{center}
\includegraphics[scale=0.75,  bb= 220 275 400 770]{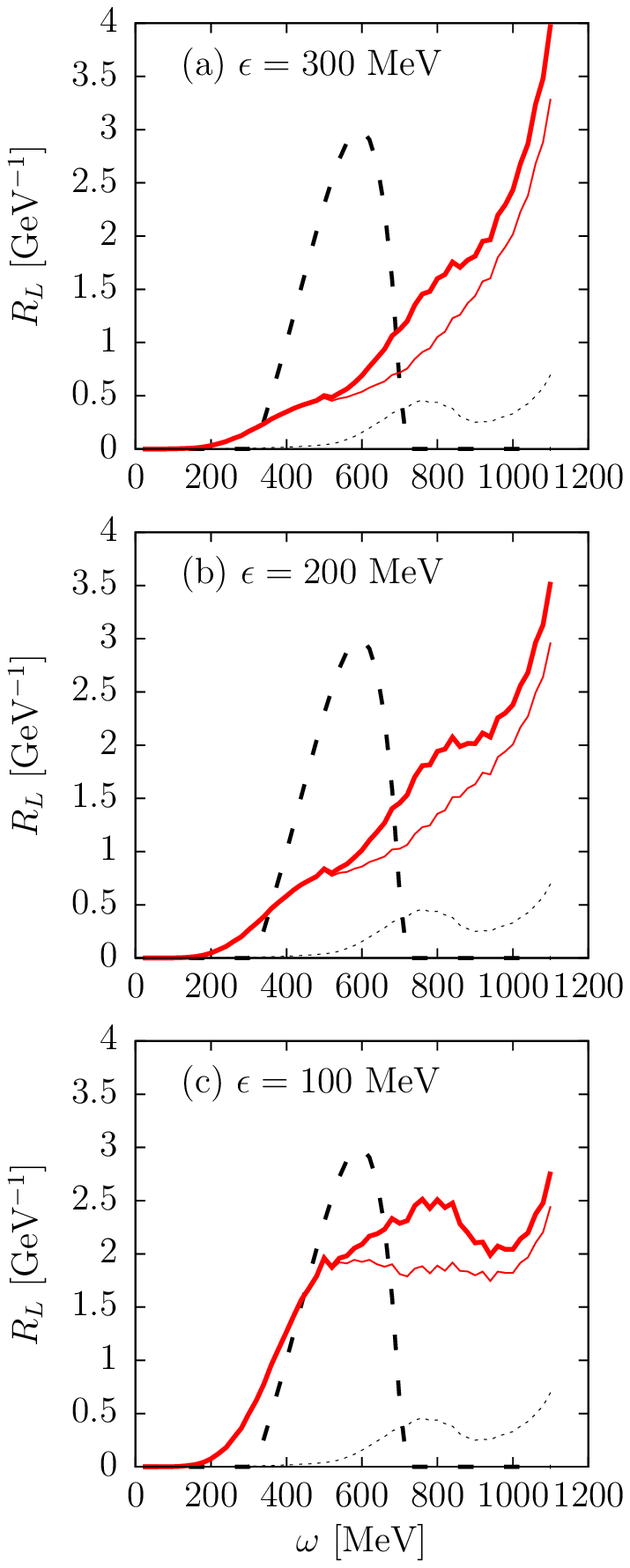}
\caption{ As for Fig. 3, but now for $R_L$ at $q=1140$ MeV/c.}
\label{Fig6}
\end{center}
\end{figure}

To make contact with previous work, we apply our model to compute
the 2p-2h longitudinal and transverse response functions for the
nucleus $^{56}$Fe, and for momenta $q=550$ and $1140$ MeV/c. The
results are presented in Figs.~3--6, where the separate
contributions of the correlation and MEC currents to the 2p-2h
responses are also shown. The 1p-1h responses produced by the
one-body (OB) current in the relativistic Fermi gas (RFG) without
interaction are also shown.

A critical input for our model is the value of the parameter
$\epsilon$ in the nucleon propagator, introduced to cure the
divergence of the double pole.  To see how the responses for the
correlation contribution depend on $\epsilon$ we show results for
three different values: $\epsilon=100, 200$ and 300 MeV. For
$\epsilon=100$ MeV, the correlation 2p-2h contribution presents a
shape with a maximum in the region of the quasielastic peak, but
with a long tail extended to high transferred energies. The maximum
is reminiscent of the pole structure of the nucleon propagator, and
therefore a resonance appears for kinematics corresponding to the
quasielastic condition in Eq.~(\ref{quasielastic}). A shift to
higher energies (of the order of $\sim 40$ MeV) is seen in the case
of $q=550$ MeV/c (Figs.~3,~5). Indeed for this value of $q$ the
phase space for two-particle emission causes a suppression of the
low-energy side of the response function.

The resonant structure produced by the 2p-2h correlation contribution
diminishes significantly with increasing values of the parameter
$\epsilon$.  Notice that for $\epsilon\geq 200$ MeV there is no maximum
located at the QE peak.

For an even lower value of the escape width, say $\epsilon = 50$
MeV, the magnitude of the resonant peak is of the same size as the
OB response function. This correction coming from 2p-2h states is
obviously too large to be compatible with experimental data that are
already of the order of the 1p-1h response at the region of the QE
peak. It should be mentioned that, although the 2p-2h contribution
should be added to the 1p-1h one, the latter should be first
corrected for final-state interaction (FSI) contributions not
included in the bare RFG results shown in the figures. In fact FSI
contribute importantly to one-nucleon emission through the coupling
of 1p-1h to 2p-2h states in the final nucleus~\cite{Smi88}. These
processes involve, in particular, two-pion exchange, and are
therefore of the same order as the 2p-2h response in the
perturbative series, since it is the square of one-pion exchange
matrix element. The inclusion of such contributions is out of the
scope of the present study.

The dependence of the correlation responses on the parameter
$\epsilon$ is better appreciated in Figs.~7 and 8, where we show its
contribution for the three chosen values of $\epsilon$ in the same
plot. In the QE region the height of the responses approximately
reduces to one half when $\epsilon$ doubles. This behavior follows
because of the leading $1/\epsilon$ dependence in
Eq.~(\ref{expansion}), coming from the pole in the propagator.  For
high $\omega$ the results are more similar and they are almost
independent of $\epsilon$ in the high-energy tail. In this case,
{\it i.e.} large $\omega$, there is no pole in the integrand and the
contribution from the propagator is less sensitive to the precise
value of $\epsilon$.

\begin{figure}[th]
\begin{center}
\includegraphics[scale=0.75,  bb= 220 490 400 770]{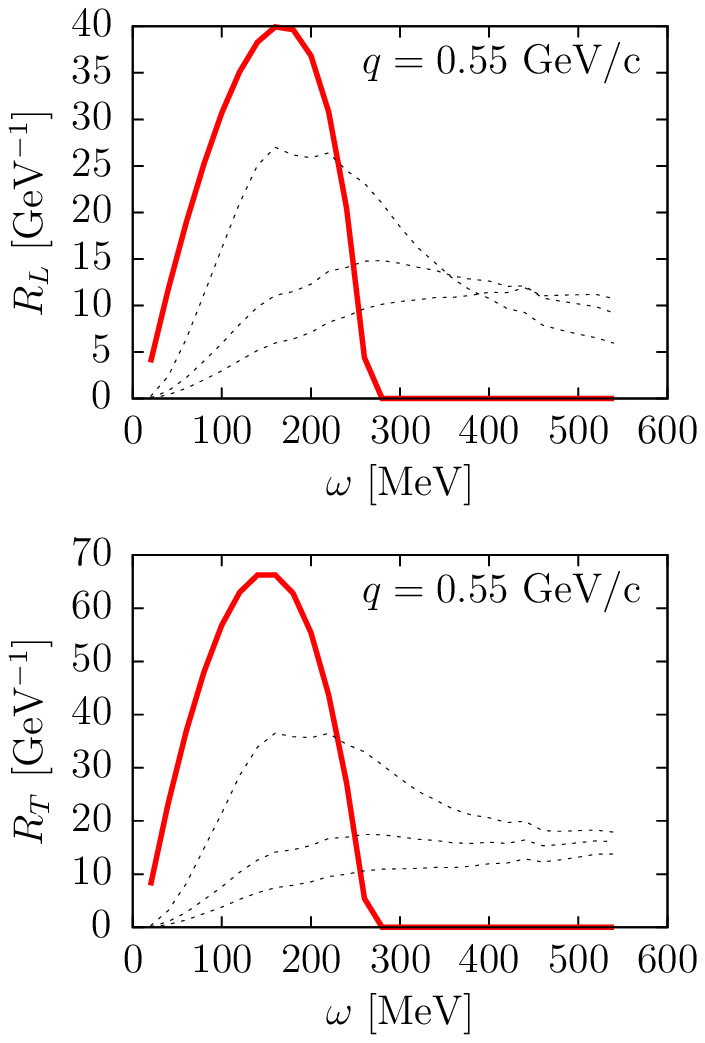}
\caption{
2p-2h correlation contribution to the
L and T responses of $^{56}$Fe for $q=550$ MeV/c.
Three values of the parameter $\epsilon$ are shown.
With dotted lines from up to down, $\epsilon=100,200,300$, respectively.
Solid lines: RFG
one-body responses.
}
\label{Fig7}
\end{center}
\end{figure}

\begin{figure}[th]
\begin{center}
\includegraphics[scale=0.75,  bb= 220 490 400 770]{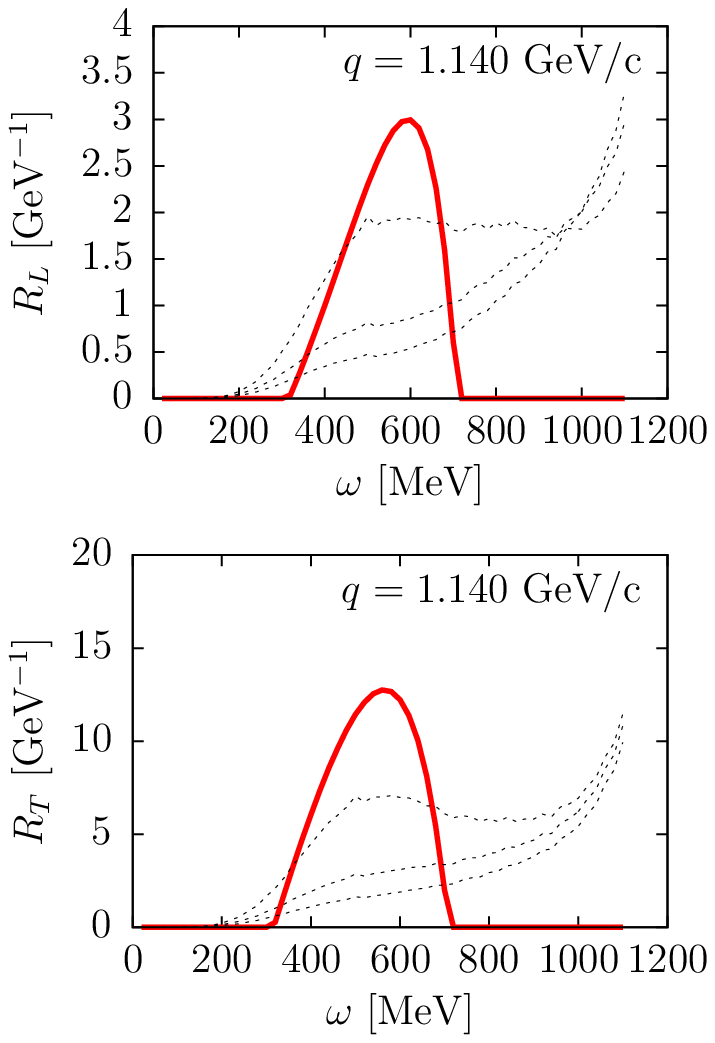}
\caption{ As for Fig. 7, but now at $q=1140$ MeV/c. } \label{Fig8}
\end{center}
\end{figure}

Let us return now to Figs.~3--6, where the MEC separate contribution
is also shown.
The transverse response (Figs.~3,~4) has a large peak with a maximum
around $\omega= (m_\Delta^2+q^2)^{1/2}-m_N$, that comes from the
$\Delta$ propagator appearing in the $\Delta$-current.  It has the
same resonant structure as the correlation current, but located in
the region of the $\Delta$ peak, where the real pion emission cross
section has a maximum. We do not include the pion emission channel
in our calculation. Both channels should be summed up to obtain the
total inclusive $(e,e')$ cross section.

The $\Delta$ peak is very small in the longitudinal response
presented in Figs.~5 and 6. This is consistent with the predominant
transverse character of the $\Delta$ current, hence providing a
small contribution to the longitudinal channel. For $q=550$ MeV the
MEC 2p-2h contribution is large (small) in the T (L) response.
However, for $q=1140$ MeV (Figs.~4,~6) we find a larger effect in
$R_L$ coming from the MEC seagull and pionic at large energy
transfer. Indeed in a non-relativistic expansion in powers of
$q/m_N$ the time component of the MEC is of higher order than the
transverse one. However, for $q=1140$ MeV, $q/m_N$ is larger than
one, and the relative L component of the MEC, compared to the T one,
starts to increase.

\begin{figure}
\begin{center}
\includegraphics[scale=0.75,  bb= 220 275 400 790]{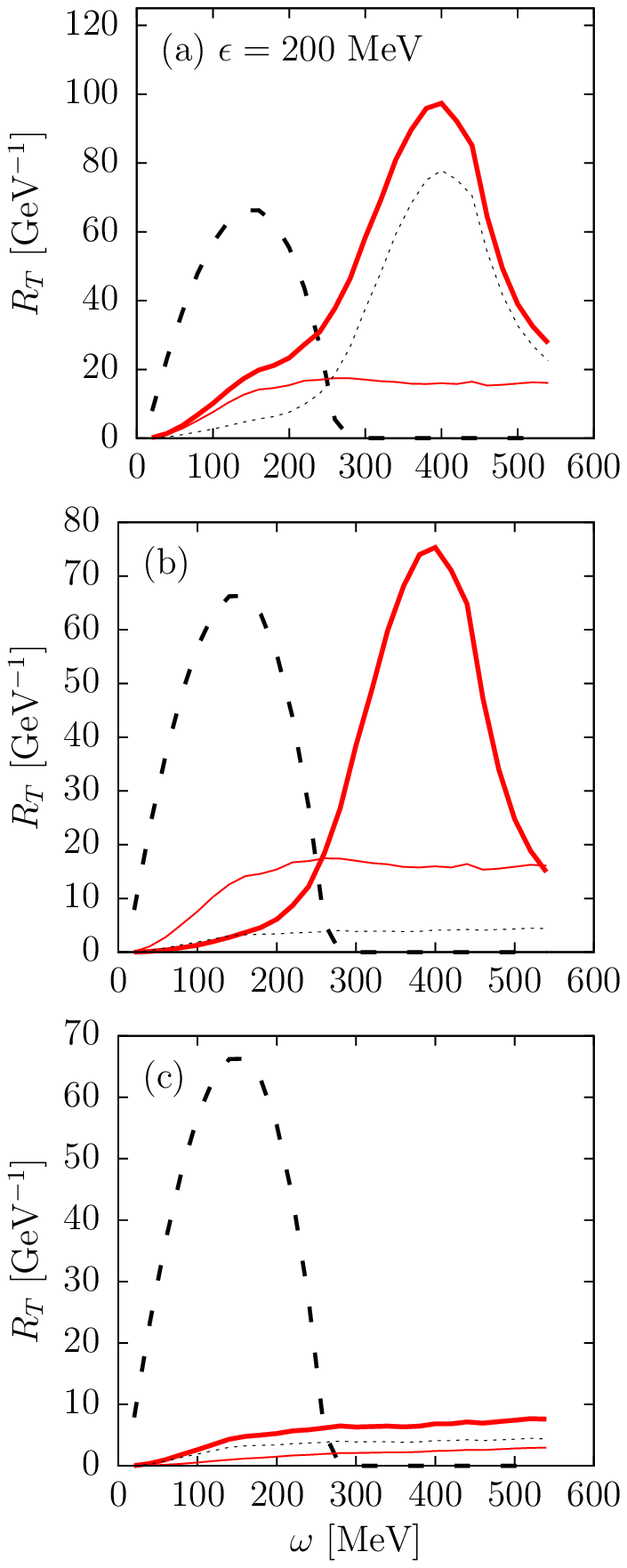}
\caption{
Contributions to the transverse response of $^{56}$Fe for $q=550$ MeV/c.
The dashed lines are the
1p-1h response with OB current only. The rest of the lines are 2p-2h
contributions. (a) Upper panel: Thin solid: Correlation only.
Dotted: MEC only. Thick solid: total. (b) Middle panel:  Thin solid:
Correlation only. Dotted: seagull+pionic only. Thick solid: $\Delta$
only. (c) Bottom panel.  Thin solid: pion-in-flight only. Dotted:
seagull+pionic only. Thick solid: $seagull$ only.
} \label{Fig9}
\end{center}
\end{figure}

\begin{figure}
\begin{center}
\includegraphics[scale=0.75,  bb= 220 275 400 770]{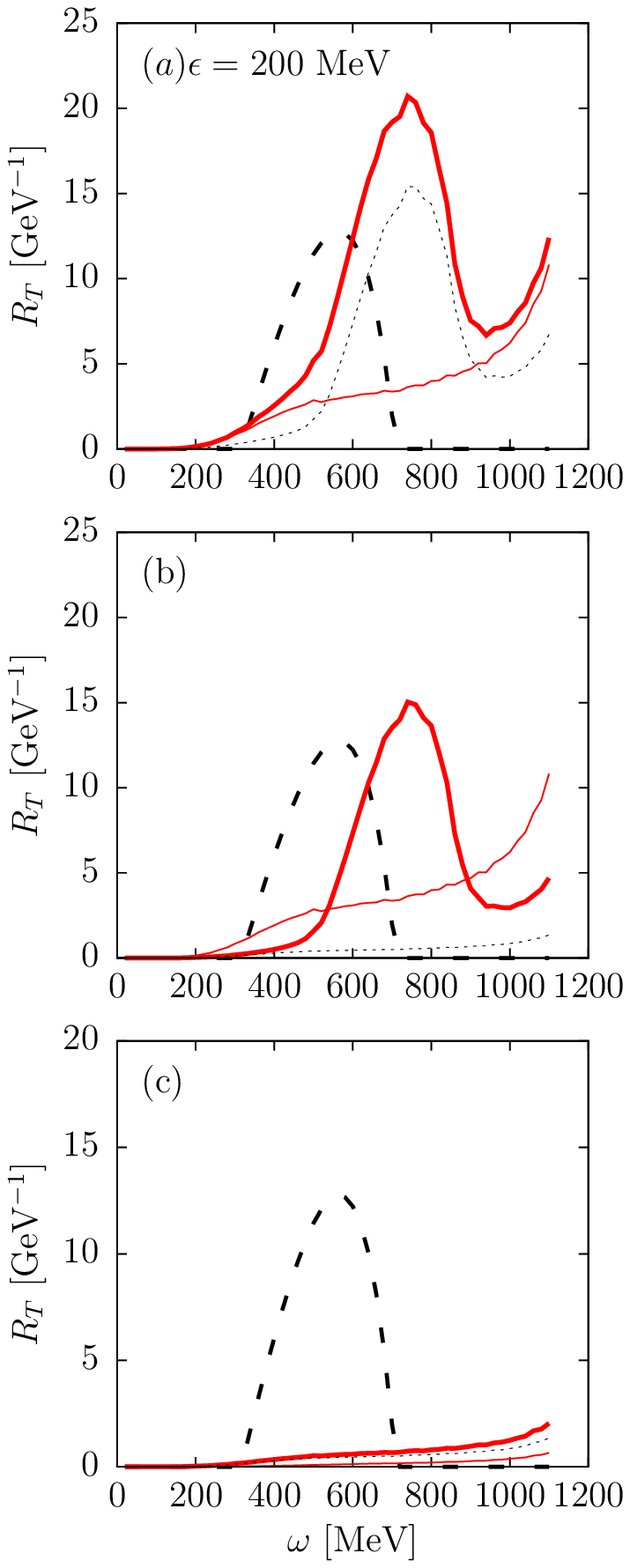}
\caption{ As for Fig. 9, but now at $q=1140$ MeV/c.
} \label{Fig10}
\end{center}
\end{figure}

In the case of the correlation current, we observe that its
contribution, compared with the OB responses, is similar in the T
and L channels. Note that in the correlation current (Fig.~2) the
photon couples directly to a nucleon with the same interaction
vertex $\Gamma^\mu$ as the OB current.  The other side of the
diagram with a pion coupled to a second nucleon is independent of
the particular component of the current.


\begin{figure}
\begin{center}
\includegraphics[scale=0.75,  bb= 220 425 400 770]{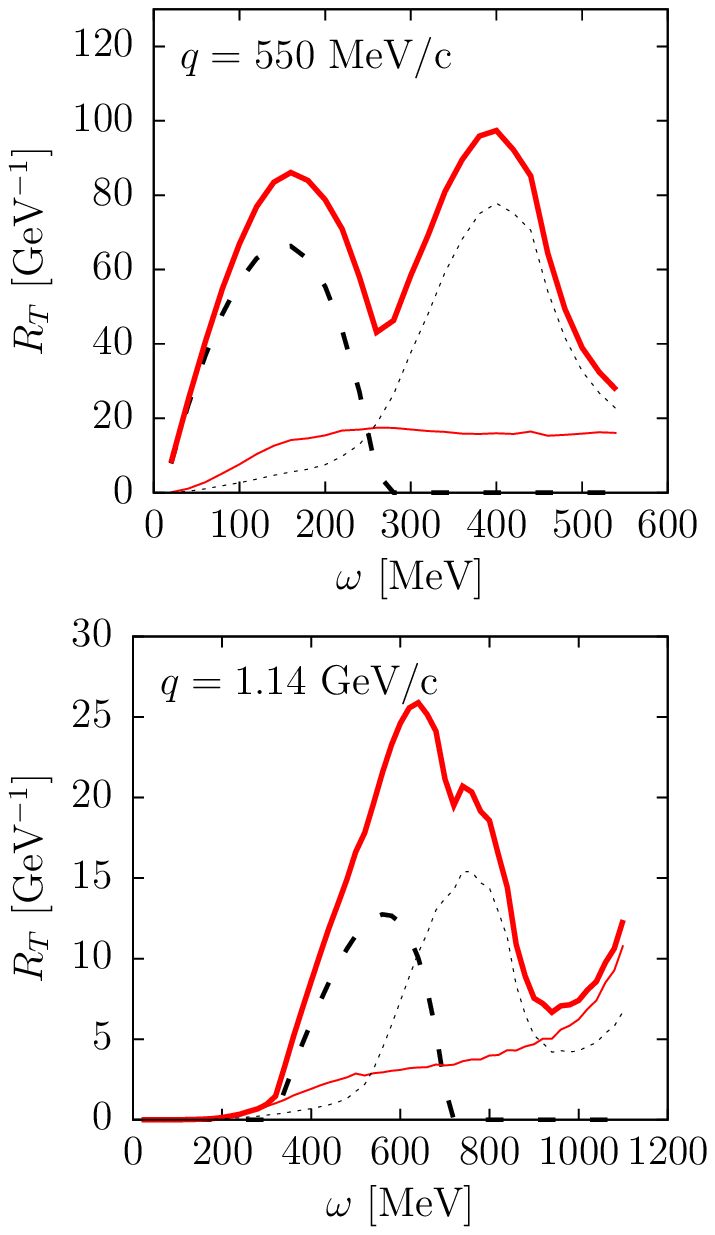}
\caption{ Transverse response of $^{56}$Fe at $q=550$ and 1140
MeV/c. Thin solid: Correlation only for $\epsilon=200$ MeV. Dotted:
MEC only. Thick solid: total one- plus two- body responses. Dashed:
RFG
1p-1h response with OB current only.
}
\label{Fig11}
\end{center}
\end{figure}

The separate effects of the different currents contributing to the
2p-2h transverse responses are shown in Figs.~9,~10.
As shown, the seagull plus pionic
(SPP) currents alone give a small effect compared with the
contributions from the $\Delta$ and correlations. In fact, for
$\epsilon=200$ MeV the correlation response is much larger (by a
factor 2 or 3) than the SPP response function (middle panels in
Figs.~9,~10). We also observe that the separate seagull contribution
is larger in magnitude than the pionic one, which is negligible for
$q=1140$ MeV/c. Note that the two currents interfere destructively
and partially cancel when both are considered in the SPP responses
(bottom panels).

\begin{figure}
\begin{center}
\includegraphics[scale=0.75,  bb= 220 485 400 770]{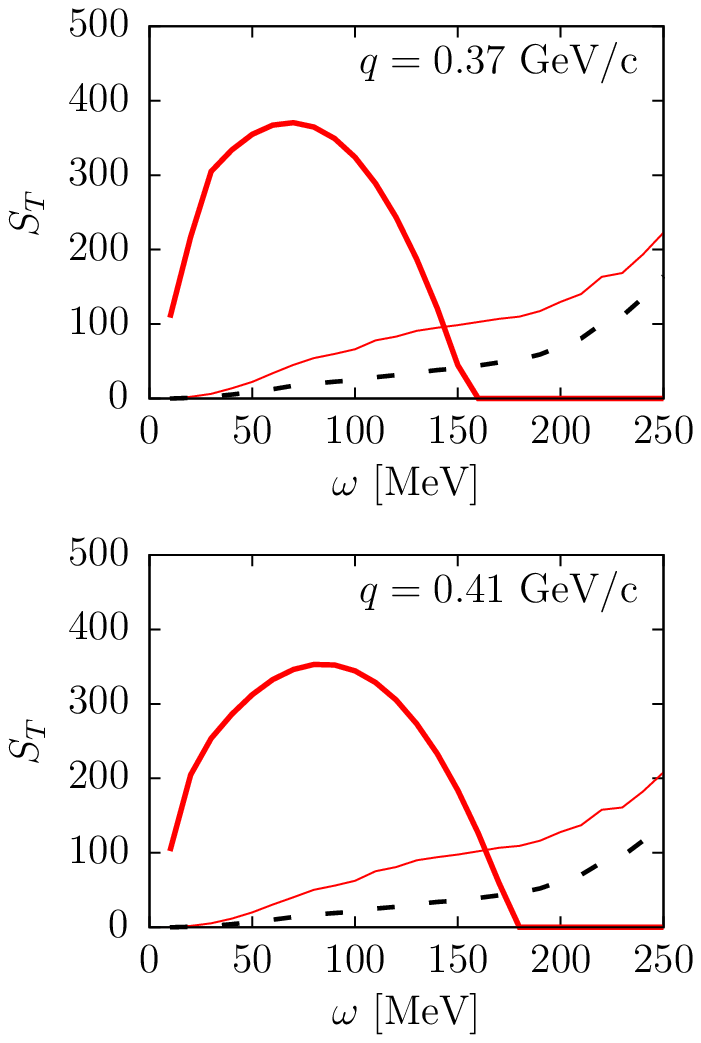}
\caption{Transverse structure function $S_T$ of $^{56}$Fe at $q=370$
and 410 MeV/c. The parameter $\epsilon=200$ MeV. To compare with
Figs. 11 and 12 of \cite{Alb84},
 $S_T$ is defined
as \cite{McC80}
$S_T=\frac{M_A}{4\pi}R_T$.
Thick lines: RFG 1p-1h results.
Dashed lines: 2p-2h, MEC only.
Thin solid lines: 2p-2h total, MEC plus correlations.
}
\label{Fig12}
\end{center}
\end{figure}

In Fig.~11 we show the transverse response obtained by adding the
total 2p-2h contributions to the OB response.
A word of caution should be raised
when analyzing these results. First, we have not added the
correlation nor MEC corrections to the 1p-1h response. Moreover, the
two-pion-exchange interaction generates  self-energy corrections to
the OB current that lead to interference effects of the same order
in the expansion as the corrections included here. As an example,
FSI are known to redistribute the strength of the responses,
producing a hardening, a reduction of the maximum and an increase of
the high-energy tail~\cite{Ama07}. Recently also a large effect from
both MEC and FSI has been found in the 1p-1h channel for high
momentum transfer~\cite{Ama10}, which should be added to the present
results. Finally, the process of real pion emission (not included
here) gives also a contribution in the transverse response located
mainly the region of the delta peak.

\begin{figure}
\begin{center}
\includegraphics[scale=0.75,  bb= 220 485 400 770]{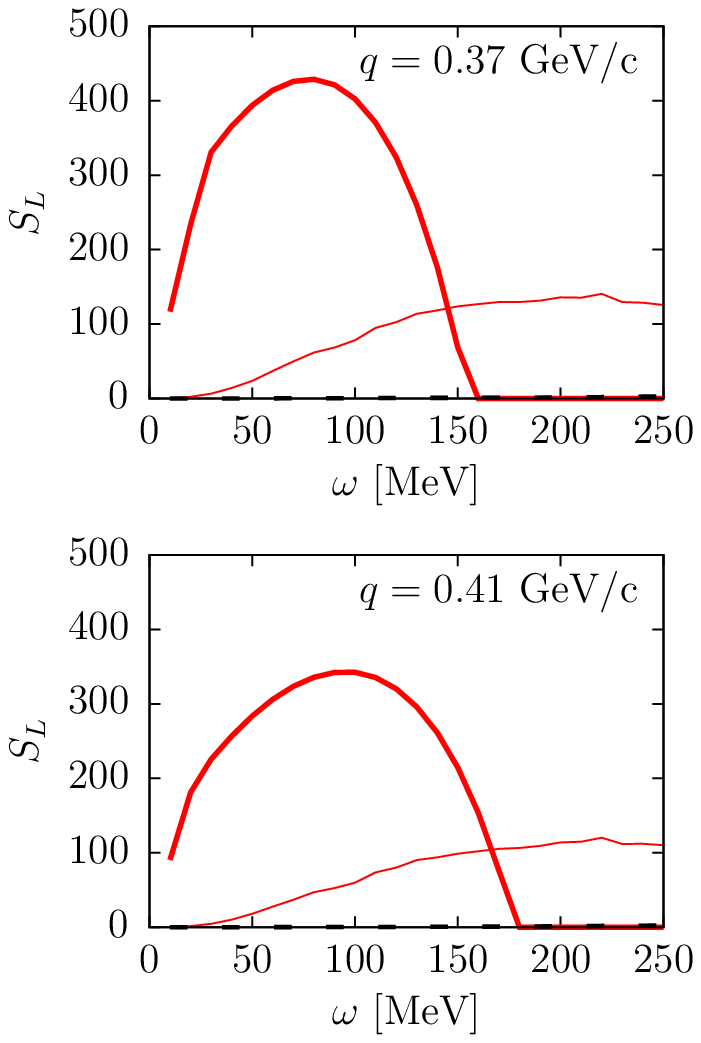}
\caption{ As for Fig. 12, but now for the longitudinal structure
function $S_L=\frac{M_A}{4\pi}R_L$. } \label{Fig13}
\end{center}
\end{figure}

So far we have presented results for intermediate to high momentum
transfer.  Results for lower values of $q=370$ and 410 MeV/c are
shown in Figs.~12 and 13 for the T and L response functions.  This
allows us to compare the present results with previous
non-relativistic calculations~\cite{Alb84}.  In Fig. 12 the
structure function $S_{T}= \frac{M_A}{4 \pi} R_{T}$ is presented, to
allow a direct comparison with Figs. 11 and 12 of \cite{Alb84}.
The separate MEC and correlation contributions to the 2p-2h T
response shown in Fig.~12 are similar to the ones presented in
\cite{Alb84}. The MEC produces a tail above the QE peak that
increases with the energy transfer. The presence of correlations
lead to an additional, significant raise of the tail. Note that our
correlation results are obtained for $\epsilon=200$ MeV. In
\cite{Alb84} another prescription to deal with the nucleon pole was
adopted. From our results we conclude that both prescriptions are
compatible numerically. The OB response of \cite{Alb84} included RPA
correlations producing a reduction and hardening of the OB response.
The 2p-2h longitudinal responses
were not computed in \cite{Alb84}, since the time components of the
MEC are of higher order in the non-relativistic reduction and hence,
they were expected to be very small. However, our prediction for the
correlation 2p-2h contribution in the L response, presented in
Fig.~13, shows a similar effect as in the T response, {\it i.e.,} a
tail also appears for high energy transfer in the L response coming
from correlations. Contrary to the T channel, MEC give no
contribution in the L response.

\begin{figure}
\begin{center}
\includegraphics[scale=0.75,  bb= 220 275 400 770]{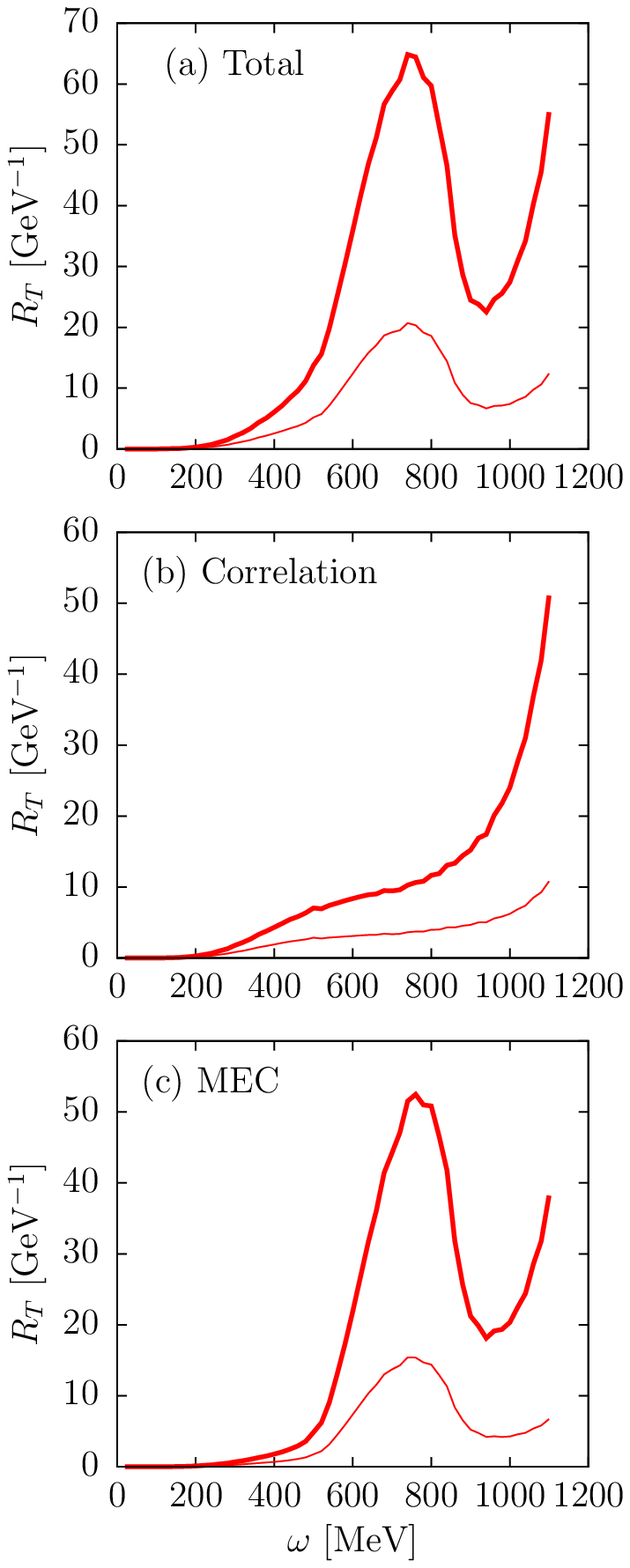}
\caption{ 2p-2h T response of $^{56}$Fe at $q=1140$ MeV/c. Thin
lines with  $\pi NN$ form factor. Thick lines without form factor
($F_{\pi NN}=1$). (a)Total, (b)  Correlations only, (c) MEC only. }
\label{Fig14}
\end{center}
\end{figure}

Since the 2p-2h excitation is produced in this work by one-pion
exchange, the results are strongly dependent on the details of this
particular interaction. This is illustrated in Fig.~14 where we show
how the results depend on the strong $\pi NN$ form factor for
$q=1140$ MeV/c. The results without a form factor, {\it i.e.,} with
$F_{\pi NN}=1$, are about three times as large as the results with the
form factor. This is different from the findings at low momentum
transfer~\cite{Ama94}, where the pion form factor can be safely
ignored.

\begin{figure}
\begin{center}
\includegraphics[scale=0.75,  bb= 220 425 400 770]{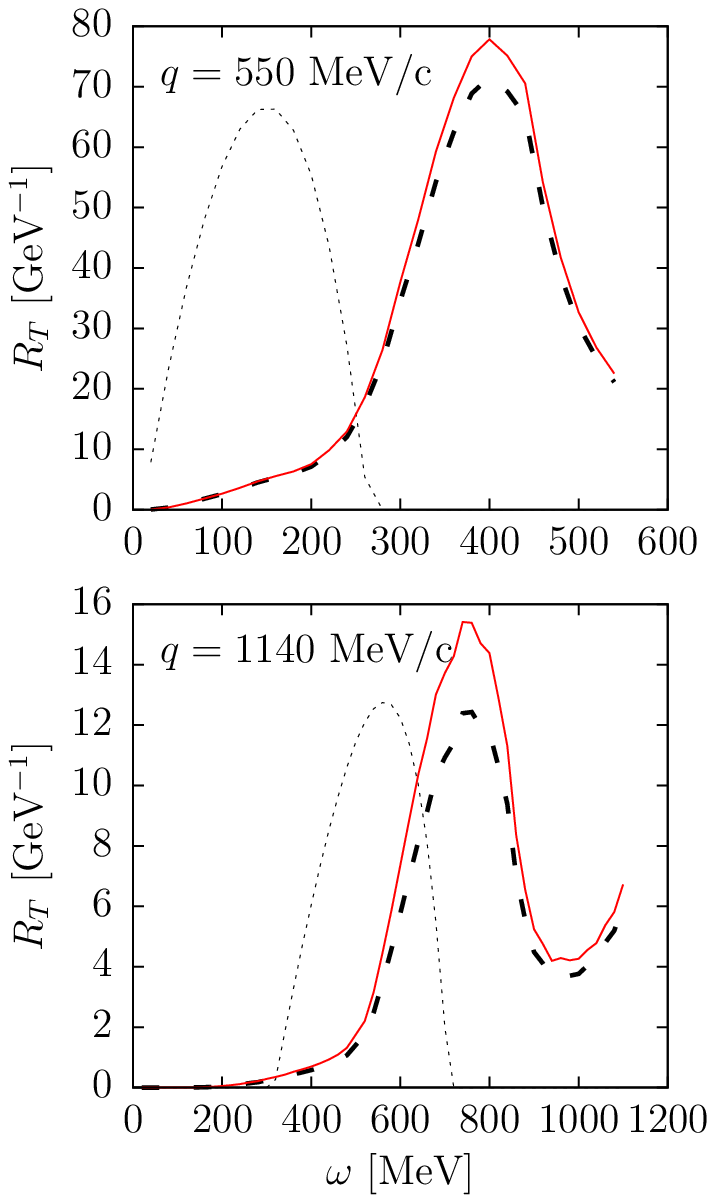}
\caption{2p2-h MEC only contribution to the transverse response of
$^{56}$Fe. Solid lines: computed with the $\Delta$ form factors used
by~\cite{Ama03}. Dashed lines: computed with the $\Delta$ form
factors used in~\cite{DePace03}. Dotted: RFG OB response. }
\label{Fig15}
\end{center}
\end{figure}

Another issue is the dependence of the results on the $\Delta$ form
factors used in this work, both the electromagnetic and the strong
ones, which are somewhat different from the parametrization used in
the 2p-2h MEC calculation of~\cite{DePace03}. Calculations done with
both sets of parameters are compared in Fig.~15. Our calculation
gives a larger contribution for the T response than the one
of~\cite{DePace03}. Hence the use of the same form factors reduces
the discrepancy between the two calculations. Some of the remaining
differences could be linked to other details of the models, in
particular to the different Lagrangian chosen for $\Delta$
electroexcitation.   We should note that the two models are fully
independent. While all the spin sums are performed analytically
in~\cite{DePace03} resulting in thousands of terms to be numerically
integrated, in this work we first compute the spin matrix elements
of the current and later we evaluate the squares and perform the
sums numerically.

Before concluding, we would like to stress that the 2p-2h responses
in the present model are crucially dependent on details of the pion
interaction. A critical ingredient of the model is the value of the
parameter $\epsilon$, identified with an escape width of a
high-energy nucleon from the nucleus. We have proven that a value
$\epsilon\sim 200$ MeV leads to results in agreement with the
previous calculation of~\cite{Alb84}. This parameter $\epsilon$ is
different from the usual interaction width of particle states,
usually associated with matrix elements of the phenomenological
imaginary optical potential derived from elastic scattering
data~\cite{Smi88}. It has also been computed in nuclear matter in a
semi-phenomenological approach~\cite{Fer92}. The resulting width for
100 MeV nucleons is of the order of 10 MeV, which is too small to
give reasonable results in our calculation. This is due to the
$1/\epsilon$ behavior of the 2p-2h response divergence in the QE
region, where the pole is being hit.

Due to this divergent behaviour, for $\epsilon=5$ MeV the results
would be almost one order of magnitude larger than the OB responses at
the maximum.  We have checked that the the $1/\epsilon$ term in the
forward diagrams is the main contribution to the 2p-2h correlations in
the QE region for $\epsilon > 20$ MeV.

The importance of correlations, for the same value of $\epsilon$,
increases with the nuclear mass. We have checked that for the case
of $^{12}$C where the sizes of the correlation responses, relative
to the OB, are about 20\% smaller than for $^{56}$Fe.  This is what
one would expect, since the number of correlated pairs increases
with $A(A-1)/2$.  Moreover, since the estimated value of $\epsilon$
depends on the nuclear radius, Eq.~(\ref{epsilon}) indicates that
one should use larger $\epsilon$-values for lighter nuclei, which in
turn would reduce even more the size of the correlation responses.
Thus we expect an important $A$-dependence of correlations on the
nuclear responses coming from the $A$-dependence of the escape width
$\epsilon$. A more deailed study of this issue will be presented in
forthcoming work.

\section{Conclusions}

In this work we have presented a fully relativistic model of
inclusive two-particle emission reactions induced by electrons.
Starting with the free relativistic Fermi gas we have considered all
Feynman diagrams in a perturbative expansion of the scattering
amplitude with one photon and one pion exchange producing 2p-2h
excitations.  Those diagrams can be classified in two sets, namely
MEC and correlation currents.  In the latter there is a nucleon
propagator that can be put on shell giving a double pole from
$(p_0-E_{\np}+i\epsilon)^{-2}$ when taking the square of the current
matrix element.  The corresponding 2p-2h response function diverges
as $1/\epsilon$ when $\epsilon\rightarrow 0$ plus additional
$\ln\epsilon$ terms. Giving a physical meaning to $\epsilon$ as the
escape width of the nucleus, namely, twice the inverse of the
nucleon propagation time, the fact that the corresponding response
is infinite is related to the infinite extension of the Fermi gas.
Using a finite value of $\epsilon$ we account for the finite size of
the nucleus, hence getting a finite result. Having no way to compute
$\epsilon$ in a Fermi gas, we take it as a parameter. Estimating in
a crude way a value around $\sim 200$ MeV, we have made an
exploratory study of the results as a function of $\epsilon$. The
correlation effects decrease with increasing $\epsilon$. Our
analysis shows that the assumption $\epsilon\sim 200-300$ MeV is not
unreasonable, whereas for  smaller $\epsilon$-values the correlation
contribution increases significantly in the QE region.

Within this framework we have studied the properties and effects of
the different 2p-2h contributions and other ingredients of the model
on the transverse and longitudinal response functions of $^{56}$Fe
for intermediate to high momentum transfer.  The MEC
give rise to a wide peak in the region of the $\Delta$ resonance
that dominates the T response. In the L channel the MEC are small
for low momentum transfer, but they importantly increase for high
momentum above the QE peak where their contribution is of the same
size as the OB longitudinal response. Concerning the correlations,
they add to the MEC in the high-energy tail and are of the same
order of magnitude. The contribution of the correlations is similar
in the L and T responses.

The main goal of this paper has been to study the effect of 2p-2h
pion correlations in the L and T response, analyzing the properties
of these effects as a function of a single parameter $\epsilon$. In
future work we plan to investigate more physically founded ways to
``fine tune'' this parameter, including its dependence on kinematics
and nuclear species. Finite-size calculations in conjunction with
the use of semi-phenomenological fits of the nucleon spreading width
or fits to existing $(e,e')$ data will also be explored.

\section*{Acknowledgments}
JEA thanks E. Ruiz-Arriola for useful discussions. This work was
partially supported by DGI (Spain): FIS2008-01143,
FPA2006-13807-C02-01, FIS2008-04189, FPA2007-62216, by the Junta de
Andaluc\'{\i}a, by the INFN-MEC collaboration agreement, projects
FPA2008-03770-E-INFN, ACI2009-1053, the Spanish Consolider-Ingenio
2000 programmed CPAN (CSD2007-00042), and part (TWD) by U.S.
Department of Energy under cooperative agreement DE-FC02-94ER40818.


\appendix
\section{Isospin matrix elements}
In the model used in this work we compute explicitly the isospin
matrix elements of the current operator in the different channels PP (two protons),
NN (two neutrons) and PN (proton-neutron) emission.

\subsection{PN channel}
We first consider the channel in which we eject a PN pair. In this
case there is no symmetry in the wave function and we assume that
the first hole is a proton and the second is a neutron, {\em  i.e.},
the initial isospin wave function is $|PN\rangle$. The final state
can be $|PN\rangle$ or $|NP\rangle$ depending on if there is or is
not charge exchange.

In the case of the MEC seagull and pion-in-flight, Fig. 1(a--c),
this is the only channel which contributes. The isospin operator is
\begin{equation}
U\equiv \epsilon_{3ab}\tau_a^{(1)}\tau_b^{(2)} \,,
\end{equation}
where repeated indices are meant to be summed. The relevant isospin
matrix element is obtained by operating over a PN state
\begin{equation}
\langle NP |U|PN\rangle = -2i \,.
\end{equation}

In the case of the correlation current we find four isospin
operators for the diagrams of Fig. 2, including the isospin
dependence in the single nucleon current $\Gamma^\mu$, namely
\begin{eqnarray}
\tau_a^{(1)}
\Gamma^{\mu(1)}
\tau_a^{(2)},
&&
\tau_a^{(1)}
\tau_a^{(2)}
\Gamma^{\mu(2)},
\\
\Gamma^{\mu(1)}
\tau_a^{(1)}
\tau_a^{(2)},
&&
\tau_a^{(1)}
\Gamma^{\mu(2)}
\tau_a^{(2)}
\,.
\end{eqnarray}
Operating over the initial
$|PN\rangle$
state we obtain
\begin{eqnarray}
\tau_a^{(1)}
\tau_a^{(2)}
\Gamma^{\mu(2)}
|PN\rangle
&=&
2\Gamma^{\mu N}
|NP\rangle -\Gamma^{\mu N}
|PN\rangle
\\
\tau_a^{(1)}
\Gamma^{\mu(2)}
\tau_a^{(2)}
|PN\rangle
&=&
2\Gamma^{\mu P}|NP\rangle -\Gamma^{\mu N}|PN\rangle
\\
\tau_a^{(1)}
\Gamma^{\mu(1)}
\tau_a^{(2)}
|PN\rangle
&=&
2\Gamma^{\mu P}|NP\rangle -\Gamma^{\mu P}|PN\rangle
\\
\Gamma^{\mu(1)}
\tau_a^{(1)}
\tau_a^{(2)}
|PN\rangle
&=&
2\Gamma^{\mu N}|NP\rangle -\Gamma^{\mu P}|PN\rangle
\,.
\end{eqnarray}
In the case of the $\Delta$ current, diagrams of Fig. 1 (d--g),
we find the following isospin operators
\begin{eqnarray}
T_a^{(1)}T_3^{\dagger(1)}\tau_a^{(2)} \,,
&&
T_3^{(1)}T_a^{\dagger(1)}\tau_a^{(2)} \,,
\\
\tau_a^{(1)}T_a^{(2)}T_3^{\dagger(2)} \,,
&&
\tau_a^{(1)}T_3^{(2)}T_a^{\dagger(2)} \,,
\end{eqnarray}
where $T_i$ are the $\frac{3}{2}\rightarrow \frac{1}{2}$ isospin transition operators verifying
\begin{equation}
T_i T_J^\dagger = \frac23\delta_{ij}-\frac{i}{3}\epsilon_{ijk}\tau_k \,.
\end{equation}
For instance we have
\begin{eqnarray}
T_a^{(1)} T_3^{\dagger(1)} \tau^{(2)}_a
&=&
\frac{2}{3} \tau_z^{(2)}
- \frac{i}{3} \left[\tauvec^{(1)} \times \tauvec ^{(2)} \right]_z
\label{iso}
\\
T_3^{(1)} T_a^{\dagger(1)} \tau^{(2)}_a
&=&
\frac{2}{3} \tau_z^{(2)}
+\frac{i}{3} \left[\tauvec^{(1)} \times \tauvec ^{(2)} \right]_z
\label{iso2}
\end{eqnarray}
and similarly changing $1\leftrightarrow2$.
Operating over the initial $|PN\rangle$ state we obtain
\begin{eqnarray}
T_a^{(1)}T_3^{\dagger(1)}\tau_a^{(2)} |PN\rangle
&=& -\frac23|NP\rangle-\frac23|PN\rangle
\\
T_3^{(1)}T_a^{\dagger(1)}\tau_a^{(2)} |PN\rangle
&=& \frac23|NP\rangle-\frac23|PN\rangle
\\
\tau_a^{(1)} T_a^{(2)}T_3^{\dagger(2)} |PN\rangle
&=& \frac23|NP\rangle+\frac23|PN\rangle
\\
\tau_a^{(1)}  T_3^{(2)}T_a^{\dagger(2)} |PN\rangle
&=& -\frac23|NP\rangle+\frac23|PN\rangle \,.
\end{eqnarray}

\subsection{PP Channel}

In the case of two proton emission only the $\Delta$ and correlation
diagrams contribute. In the case of the correlations, the isospin
operators over the initial $|PP\langle$ state give
\begin{equation}
\tau_a^{(1)}
\tau_a^{(2)}
\Gamma^{\mu(2)}
|PP\rangle
=
\Gamma^{\mu P} |PP\rangle
\end{equation}
and exactly the same result for the remaining three operators.

In the case of the $\Delta$ we have
\begin{equation}
T_a^{(1)}T_3^{\dagger(1)}\tau_a^{(2)} |PP\rangle
= \frac23|PP\rangle
\end{equation}
and exactly the same answer for the remaining three operators.

\subsection{NN Channel}

Once more only the $\Delta$ and correlation diagrams contribute. In
the case of the correlations, we have
\begin{equation}
\tau_a^{(1)}
\tau_a^{(2)}
\Gamma^{\mu(2)}
|NN\rangle
=
\Gamma^{\mu N} |NN\rangle
\end{equation}
and the same for the remaining three operators.

Finally, for the $\Delta$ we have
\begin{equation}
T_a^{(1)}T_3^{\dagger(1)}\tau_a^{(2)} |NN\rangle
= -\frac23|NN\rangle
\end{equation}
and  the same answer again for the remaining three operators.

\section{Integration of the energy delta function}

The 9-D integral for the 2p-2h response functions is of the type
\begin{equation}
\int d^3p'_1 d^3h_1 d^3 h_2
\delta(E_1+E_2+\omega-E'_1-E'_2)f(\nh1,\nh2,\np'_1,\np'_2) \, ,
\end{equation}
where $\np'_2=\nh_1+\nh_2+\nq-\np'_1$.  The delta function allows us
to perform one integration analytically imposing energy
conservation. Therefore, for fixed values of the two hole momenta
$\nh_1,\nh_2$ and for fixed values of the two emission angles
$\theta'_1,\phi'_1$ of particle 1, we can integrate over the
momentum $p'_1$, fixing the energy of the first particle. To this
end we change variables $p'_1\rightarrow E'=E'_1+E'_2$.  Taking into
account that both energies $E'_1$ and $E'_2$ depend on $p'_1$ to
compute the Jacobian of the transformation, we obtain
\begin{equation}
dp'_1=\frac{dE'}{\left|
\frac{p'_1}{E'_1}-\frac{\np'_2\cdot\np'_1}{E'_2p'_1} \right|} \, ,
\end{equation}
where the momentum of the final nucleon for fixed emission angles
$\theta'_1,\phi'_1$ is obtained by solving the energy conservation
equation. This is a second degree equation with two solutions given
explicitly by
\begin{equation}
p'_1=\frac{a}{b}\left(v\pm v_0\sqrt{1-\frac{bm_N^2}{a^2}}\right) \,
,
\end{equation}
where
\begin{eqnarray}
a &=& \frac12 p'{}^2 \\
b&=& E'{}^2-p'{}^2\cos^2\beta'_1 \\
v_0 &=& E' \\
v &=& p'\cos\beta'_1 \,,
\end{eqnarray}
 $E'=E_1+E_2+\omega$ is the final total energy,
$\np'=\nh_1+\nh2+\nq$ is the final total momentum and $\beta'_1$ is
the angle between $\np'_1$ and $\np'$.  To compute the integral we
add the contributions from these two solutions, corresponding to two
possible final states compatible with energy-momentum conservation.


\end{document}